JOURNAL ARTICLE

*Anh-Minh Tang,[1] Yu-Jun Cui,[2] Nathalie Barnel[3]*

# A new isotropic cell for studying the thermo-mechanical behavior of unsaturated expansive clays

**ABSTRACT:** This paper presents a new suction-temperature controlled isotropic cell that can be used to study the thermo-mechanical behavior of unsaturated expansive clays. The vapor equilibrium technique is used to control the soil suction; the temperature of the cell is controlled using a thermostat bath. The isotropic pressure is applied using a volume/pressure controller that is also used to monitor the volume change of soil specimen. Preliminary experimental results showed good performance of the cell.

[1] Research Engineer, Ecole Nationale des Ponts et Chaussees, Paris, (CERMES) – Institut Navier, 6 et 8, av. Blaise Pascal, Cite Descartes, Champs-sur-Marne, 77455 MARNE – LA - VALLEE CEDEX 2, France. Email : tang@cermes.enpc.fr.
[2] Professor, Ecole Nationale des Ponts et Chaussees, Paris, (CERMES) – Institut Navier, 6 et 8, av. Blaise Pascal, Cite Descartes, Champs-sur-Marne, 77455 MARNE – LA - VALLEE CEDEX 2, France. Email : cui@cermes.enpc.fr.
[3] Research Engineer, Electricite de France, Departement MMC, Group Couplages Physique-Mecanique, site des Renardieres, Avenue des Renardieres, Ecuelles, 77818 Moret sur Loing cedex, France.

## Introduction

Compacted expansive clay is often considered as a possible buffer material for deep nuclear waste disposal. In this context, this material is submitted to heating by the waste packages, to water infiltration from the geological barrier, and to the mechanical solicitation due to its interaction with the geological barrier (clay swelling, geological barrier deformation, etc.). Full-scale experimental tests showed that the temperature inside the engineered buffer was ranging from 20 °C to 100 °C; the relative humidity varied between 30 % and 100 %, which corresponds to suction range of 1 – 500 MPa; the mean net stress could reach 10 MPa (Huertas et al. 2000; Chijimatsu et al. 2001). In the work described here, a suction-temperature controlled isotropic cell is developed in order to study the thermo-mechanical behavior of unsaturated compacted expansive clay under these conditions.

Several thermo-mechanical tests have been performed on clays at saturated state. In these works, oedometer cell is often immersed in a thermostat bath (Towhata et al. 1993) for controlling the soil temperature. The temperature of soil specimen in triaxial cell is often controlled by a heating coil that covers the outer wall of the cell (Delage et al. 2000) or by a heater that is immersed in the confining liquid of the cell (Cekerevac et al. 2003).

Three following techniques are often used to control the suction: (1) axis translation; (2) osmotic; (3) vapor equilibrium. The axis translation technique is usually used to generate suction from 10 to 1500 kPa; the suction value can reach 14 MPa in some special cases (Romero et al.

2005). Suction generated by osmotic technique is comprised between 0 and 8.5 MPa (Cuisinier and Masrouri 2004) and may reach 12 MPa (Delage et al. 1998). However, highly compacted expansive clay becomes saturated at suction lower than 1 MPa (Villar and Lloret 2004); in addition, the measurement in a full-scale test showed that suction in clay buffer is ranging from 1 to 500 MPa (Huertas et al. 2000). Obviously, only the vapor equilibrium technique can cover the full suction range of these soils.

Study on thermo-mechanical behavior of unsaturated soils required simultaneous control of suction and temperature. The triaxial apparatus developed by Saix and Jouanna (1990) allowed the control of capillary suction from 0 to 100 kPa and a control of temperature from 25 °C to 80 °C. This apparatus was used to study the thermo-mechanical behavior of unsaturated silty sand (Saix 1991, Saix et al. 2000, Jamin et al. 2003). The axis translation technique was used to control the soil suction in oedometer (Romero et al. 1995) and in triaxial apparatus (Romero et al. 1997) under temperatures varying from 22 °C to 80 °C. As mentioned above, only the vapor equilibrium technique is able to cover the suction range from 1 to 500 MPa of clay buffer.

The aim of the work described here is to develop a mechanical device that enables the study of mechanical properties of compacted expansive soil using simultaneous control of suction (within the full range from 1 to 500 MPa) and temperature (ranging from 20 to 80 °C). An isotropic cell is developed, the vapor equilibrium technique is used to control the soil suction, and the cell is immersed in a thermostat bath for temperature control.

## Experimental set-up

The basic scheme of the new cell is presented in Figure 1. The soil specimen (10 mm high and 80 mm in diameter) is sandwiched between two ceramic porous stones that are built in two metallic plates. This type of ceramic porous stones is often used in the drainage system of standard oedometer. The metal used for the plates is X30CR13, which is a stainless steel having high rigidity (Modulus of elasticity: $215\times10^3$ MPa). Small holes (2 mm in diameter) are drilled on the lower plate. A glass cup containing a saturated saline solution is placed in the chamber below this lower plate to control the soil suction. A thermocouple is installed inside the cell to monitor the cell temperature. This temperature is considered equal to the temperature of the soil specimen during a test. A neoprene membrane of 1.2 mm thick covers the soil specimen and the metallic plates.

The experimental set-up of the suction-temperature controlled compression test is presented in Figure 2. The cell is immersed in a thermostat bath. A thermostat pump controls the temperature of this bath as well as that of the cell with a fluctuation of ±0.1 °C. A volume/pressure controller controls the confining pressure of water inside the cell. This controller is also used to monitor the volume change of the soil specimen. A cooling bath is used to balance any temperature effect on the volume change measurement of the volume/pressure controller. During tests, data as pressure and volume of water in the controller, cell temperature are recorded in computer. Using this cell, loading/unloading tests can be performed at constant suction and temperature. In addition, heating/cooling test at constant pressure can be also run on unsaturated soils.

In order to check a possible effect of the vapor circulation on the duration needed to reach the equilibrium in the test, a second isotropic cell is developed. Its basic scheme is presented in Figure 3 and the experimental set-up is presented in Figure 4. The configuration of this cell is similar to that presented in Figure 1, but the suction controlling method is modified. The lower plate is connected with an air circulation system that allows airflow through the lower porous stone. The air outlet and inlet of the cell are connected to an air circulation system to control the soil suction using vapor equilibrium technique. This air circulation is ensured by a pump. This modified cell can be used to run either loading/unloading tests under constant suction or wetting/drying tests under constant pressure, all under ambient temperature.

The technical specification of the volume/pressure controllers used in this study is presented in Table 1. These controllers are provided by GDS Instrument LTD Company. Two types of controllers are used: I and II. Controller I provides a better resolution in pressure (2 kPa), and is used for tests at low pressures (p ≤ 3 MPa), whilst Controller II provides a less exact pressure resolution (15 kPa) and is used for higher pressures. The technical specification shows equally that the measurement of water volume change using these controllers can be affected by temperature, pressure and measured volume. Calibration tests are therefore necessary to minimize these errors.

Table 1. Technical specification of volume/pressure controllers

| Specification | Controller I | Controller II |
|---|---|---|
| Pressure ranges | 3 MPa | 64 MPa |
| Volumetric capacity | 200 000 mm$^3$ | 200 000 mm$^3$ |
| Pressure/volume resolution | 1 kPa / 1 mm$^3$ | 15 kPa / 1 mm$^3$ |
| Pressure accuracy | 4.5 kPa | - Characteristic error: 64 kPa + 0.05 % measured value |
| | | - Environmental error: 0.025 %/°C measured value |
| Volume accuracy | 0.25 % measured value | - Characteristic error: 0.25 % measured value |
| | | - Environmental error: 0.02 %/°C + 0.2 %/MPa volume of water in cylinder |

**Material and experimental procedures**

The studied material is MX80 clay, one of the reference materials proposed for the engineered barrier in deep nuclear disposal. The identification parameters of MX80 clay are presented in Table 2. The clay contains 82 % montmorillonite; its plasticity index is quite high (474 %); its specific surface is important (S = 562 m$^2$/g); its main exchangeable cations are Na$^+$ and Ca$^{2+}$. MX80 is classified as very highly plastic clay.

Table 2. Identification parameters of MX80 clay.

| Property | MX80 |
| --- | --- |
| Montmorillonite content (%) | 82[a] |
| Particle density (Mg/m$^3$) | 2.76[b] |
| Liquid limit (%) | 520[a] |
| Plastic limit (%) | 46[a] |
| Plasticity index | 474[a] |
| Cation exchange capacity (meq./100g) | 76[b] |
| Exchange capacity of Na$^+$ (meq./100g) | 62.4[b] |
| Exchange capacity of Ca$^{2+}$ (meq./100g) | 7.4[b] |
| Specific surface area, S (m$^2$/g) | 562[b] |

[a] Sauzéat et al. (2000); [b] Madsen (1998)

Prior to utilization, the clay was sieved at 2 mm and dried at 44 % relative humidity (at 20 °C, that corresponds to 110 MPa suction) using the vapor equilibrium technique. When the equilibrium of soil weight was reached, the water content of soil was 9.5 %. At this state, clay powder was compacted in an isotropic cell under 40 MPa pressure. The dry density of soil sample after compaction was 1.80 Mg/m$^3$. The compacted sample, 90 mm in diameter and 120 mm high, was then machined to obtain smaller samples: 80 mm in diameter and 10 mm high. The small samples were then placed inside a hermetically sealed box at 76 % relative humidity (at 20 °C, that corresponds to 39 MPa suction). They were weighed every 3 days until weight stabilization. After this wetting, the dry density was 1.45 Mg/m$^3$ and the water content was 17.2 %.

Saturated sodium chloride (NaCl) solution was used to control the soil suction in the cell. Tang and Cui (2005) measured the relative humidity generated by saturated saline solutions at temperatures varying from 20 to 60 °C and noted that the suction given by sodium chloride solution was practically temperature independent. This solution imposed a suction of 39 MPa to the soil at temperature ranging from 20 to 60 °C. This value of suction is equally the initial suction in soil specimen in the work described here.

Three tests were performed following the program showed in Table 3. Tests T01 and T03 were carried out in the suction-temperature controlled isotropic cell (Figure 1). Test T02 was performed in the isotropic cell that has suction control by vapor circulation (Figure 3). In test T01, after a heating/cooling cycle of 25 − 70 − 25 °C, a loading/unloading cycle was applied while the temperature was maintained constant at 25 °C. The same loading/unloading cycle was applied in test T02 at ambient temperature. For test T03, the loading/unloading cycle was applied at 60 °C temperature.

Table 3. Tests program.

| Test | Procedure |
| --- | --- |
| T01 | Heating/cooling cycle: 25 − 70 − 25 °C (Controller I, p = 0.1 MPa); Loading/unloading cycle: 0.1 − 50 − 1 MPa (Controller II, T = 25 °C). |
| T02 | Loading/unloading cycle: 0.1 − 50 − 2 MPa (Controller II, T = 20 °C). |
| T03 | Heating: 25 − 60 °C (Controller I, p = 0.1 MPa); Loading/unloading cycle: 0.1 − 50 − 0.2 MPa (Controller II, T = 60 °C). |

**Experimental results**

*Test T01*

After initial equilibrium, the cell was first heated sequentially from 25 to 30, 40, 50, 60 and 70 °C. The water volume change and the cell temperature are presented versus time in Figure 5 during this heating. It appears that the temperature inside the cell reached rapidly stabilization after each adjustment of the thermostat pump. Heating generated a dilation of water inside the cell and pushed water out, increasing the volume of water inside the volume/pressure controller. It is observed in Figure 5 that the water volume change reached stabilization just after the stabilization of cell temperature. The data for the cooling path from 70 °C to 25 °C was unfortunately not recorded.

The water volume change was plotted versus the cell temperature for the heating path in Figure 6. In order to determine the volume change of the soil specimen, the results of the calibration test is plotted in the same figure. The calibration test was performed on a metallic specimen having the same dimensions of the soil specimen. After Romero et al. (2005), the thermal dilation coefficient of compacted expansive soil is approximately $10^{-4}$ °C$^{-1}$, much higher than that of metal ($10^{-6}$ °C$^{-1}$). It can be then assumed that the thermal dilation of metal was negligible in comparison with that of soil. Therefore, in Figure 6, the soil volume change can be estimated as the difference between the two curves obtained. It is observed that the volume of soil specimen increased 500 mm$^3$ during heating from 25 °C to 70 °C.

After the thermal cycle, a loading/unloading cycle at constant temperature was applied to the soil specimen. The temperature was kept constant at 25 °C; the pressure was increased from 0.1 MPa to 0.2, 0.5, 1, 2, 5, 10, 20, 30, 50 MPa. Each load was maintained during about one week for volume stabilization. The water volume change in each loading step is presented in Figure 7. The results of calibration test performed on a metallic specimen having the same dimensions of the soil specimen are equally presented. Assuming that the volume change of the metallic sample is negligible under pressure (Modulus of elasticity of the metal used, X30CR13, is $215 \times 10^3$ MPa), the calibration curves showed in Figure 7 represent the deformation of the system (cell and tubing) under pressure.

In order to estimate the soil volume change during each loading step, the results obtained from the test and from the calibration must be used together. In the case of 1 MPa loading step, for example, the volume of water in the controller reduced 700 mm$^3$ in the calibration test. For the test on soil specimen, this value was 1000 mm$^3$ after 1 minute; it stabilized at 1500 mm$^3$ after 4 days. It can be concluded then that the volume of soil specimen decreased instantly 1000 − 700 = 300 mm$^3$ and that the final soil volume change was 1500 − 700 = 800 mm$^3$. The volumetric strain of the soil specimen can be then calculated. Villar (2000) and Marcial (2003) also observed these two types of deformation during oedometric tests at controlled suction on unsaturated compacted expansive clay. The first one corresponds to the instantaneous compaction of soil. After Delage et al. (2006), this compaction induces water redistribution inside compacted expansive soil and a possible suction decrease. The secondary deformation observed in suction controlled compression test corresponds to the suction re-equilibrium between soil suction and suction imposed by the saturated saline solution. As soil suction decreases after the compaction, imposing initial suction by the saturated saline solution corresponds to a drying process in soil, and that induces then a secondary compaction in expansive soil.

The estimation of soil volume change during each loading stage was not an easy task. On one hand, the water volume change when loading from 0.1 to 0.2 MPa and from 0.2 to 0.5 MPa was smaller than that in the calibration test. This was due to the inaccuracy of the volume/pressure controller at low pressures. In these cases, it was considered that the instantaneous strain of soil upon loading was negligible, and that the secondary strain was the difference between the final volume and the volume after 1 minute (instantaneous volume change) of the curve obtained from the test. The volume changes of soil specimen under these loads were then 50 mm$^3$ (for the loading step at 0.2 MPa) and 450 mm$^3$ (for the loading step at 0.5 MPa). On the other hand, during the loading from 2.0 to 5.0 MPa, an abrupt decrease of water volume was observed after 24 hours. This was caused by a leakage in the tubing of the confining system that was repaired afterward. Therefore, the volume change during this loading step was taken at 24 hours.

Unloading was carried out after the 50 MPa loading step. The pressure was reduced sequentially from 50 MPa to 20, 10, 5, 2 and 1 MPa. In Figure 8, water volume changes were plotted versus time for each unloading step. As the swelling of compacted bentonite under unloading was small, the difference between raw results and calibration one was so small that it is difficult to use it to estimate the soil volume change. Thus, only secondary strains were estimated and the soil volume change was considered as the difference between the water volume after 5 minutes and the final volume on the curves obtained from the test.

*Test T02*

In case of test T02, after applying the initial conditions (s = 39 MPa, p = 0.1 MPa), the soil specimen was loaded sequentially from 0.1 MPa to 0.2, 0.5, 1, 2, 5, 10, 20, 30, 50 MPa under ambient temperature (20 °C). The results obtained are presented in Figure 9, where the volume change was plotted versus time for each loading step. The phenomenon observed in the test T01 can be also observed in this test: after a pressure increase, instantaneous decrease of water volume is followed by a secondary decrease.

After loaded under 50 MPa, the soil specimen was unloaded to 20, 10, 5, and 2 MPa. The corresponding water volume changes are presented in Figure 10. At the end of the unloading increment from 50 MPa to 20 MPa, a technical problem was detected: condensed water blocking the vapor circulation was observed inside the tubing. After resolving the problem, suction control was reestablished, giving rise to an additional volume decrease as showed in Figure 10.

*Test T03*

After the application of the initial conditions (T = 25°C, s = 39 MPa, p = 0.1 MPa), the soil specimen was heated from 25 °C to 60 °C. When the volume change reached the stabilization, it was loaded sequentially from 0.1 MPa to 0.2, 0.5, 1, 2, 5, 10, 20, 50 MPa and then unloaded with the same increments until 0.1 MPa. The results obtained in term of water volume changes were similar to that of test Iso01.

**Discussion**

*Volumetric behavior under heating*

In Figure 11 the thermal volumetric strains deduced from tests T01 and T03 were plotted versus temperature. Data during heating from 40 °C to 60 °C were unfortunately not recorded. In both cases, heating induced a negative volumetric strain corresponding to a thermal dilation. The two tests gave similar dilation results, showing the satisfactory efficiency of the experimental set-up for thermal tests. A thermal dilation coefficient of $2\times10^{-4}$ °C$^{-1}$ can be calculated from these results. This value is close to that obtained by Romero et al. (2005) on an unsaturated compacted expansive clay under mechanical unconfined conditions.

*Compressibility properties*

Figure 12 presents the compression curves deduced from tests T01 and T02. In the case of test T01, it was assumed that the thermal dilation is elastic and thus the soil volume does not change after the thermal cycle. Therefore, the void ratio at initial pressure (0.1 MPa) for the two tests are similar, $e = 0.900$. A good agreement between the two curves can be observed. During loading from 0.1 MPa to 1 MPa, the values of void ratio for test T02 were slightly smaller than that for test T01. This difference may be related to the inaccuracy of Controller II (high pressure) at low-pressure range. The results in the range from 2 MPa to 10 MPa during loading and in the range from 20 MPa to 2 MPa during unloading are almost similar. The only significant difference can be found when loading from 10 MPa to 50 MPa and when unloading from 50 MPa to 20 MPa. This difference can be explained by the technical problem mentioned previously in test T02. The air circulation blockage would occur when loading from 10 MPa to 20 MPa. It is interesting to observe that when unloading at 20 MPa, the curve of test T02 came back to the position of test T01, showing the resolution of the technical problem.

For test T01, loading from 0.1 MPa to 50 MPa reduced the void ratio of soil from 0.900 to 0.538. When the soil was unloaded to 1 MPa, the void ratio was 0.572, much lower than that at 1 MPa on the loading path (0.854), showing that loading until 50 MPa induced plastic strains. The soil behavior can be then considered as elasto-plastic.

As mentioned above, test T01 was performed in a cell that has no air circulation in the system of suction control, as opposed to test T02. Similar systems of suction control without air circulation were used in oedometer tests on unsaturated soils performed by Belanteur et al. (1997), Cuisinier and Masrouri (2004). The system of suction control using air circulation was also used by several authors as Bernier et al. (1997), Cui et al. (2002), Lloret et al. (2003). The objective of this air circulation was to accelerate water exchange between soil specimen and saline solution. Nevertheless, the results of tests T01 and T02 did not show any effect of the circulation system (see Figure 7 to 10). The explanation may be that the change in soil suction during loading/unloading steps was not significant.

*Temperature effect*

Figure 13 gathered the results from tests T02 (20 °C) and T03 (60 °C). After heating from 25 °C to 60 °C, the void ratio in test T03 increased from 0.900 to 0.912. As the information when heating from 40 °C to 60 °C in this test was lost, this void ratio value was estimated by assuming a linear thermal dilation with a coefficient of $2 \times 10^{-4}$ °C$^{-1}$. In general, it can be observed that the curve of test T03 is similar to that of test T02. The compression index ($C_c$) and the swelling index ($C_s$) estimated from these two curves were identical. Only a decrease in preconsolidation pressure ($p_0$) was observed during heating: it decreased from 1.9 MPa (at 20 °C) to 0.8 MPa (at 60 °C). As mentioned above, the both tests were performed at 39 MPa suction, the decrease of preconsolidation pressure due to heating observed here can be then explained by the temperature independence of soil strength at unsaturated state.

The independence of the swelling index and the compression index observed in the work described here is in agreement with the tests performed on saturated clays by Cekerevac and Laloui (2004). Other works on unsaturated silty sand performed by Recordon (1993) and Saix et al. (2000) showed also this independence. The decrease of the preconsolidation pressure with temperature increase is also consistent with the results of Sultan et al. (2002) on saturated Boom clay and that of Romero et al. (2003) on unsaturated compacted Boom clay.

*Cell performance*

The accuracy of the experimental set-up involves the pressure control and volume measurement by the volume/pressure controllers, the suction control, the temperature control as well as the deformation behavior of the cell and the tubing. Table 1 shows that the controllers used offer satisfactory pressure and volume accuracy. The utilization of thermostat bath for temperature control was found to be a good solution since it permits to control the fluctuation within ±0.1 °C. Placing the whole cell in water at constant temperature has been shown to be an efficient method to minimize the effect of room temperature changes. As far as the suction control is concerned, the accuracy is closely related to the temperature fluctuation. With a fluctuation of ±0.1 °C, Tang and Cui (2005) showed that, using saturated NaCl solution, the suction can be controlled at 39 ± 0.05 MPa. To verify the effect of suction control method on the equilibrium time in mechanical test, a second cell was developed using a vapor circulation to control the soil suction. Comparison between results obtained from these two cells shows that the vapor circulation did not affect the equilibrium time significantly. Moreover, the results of the two tests performed at ambient temperature and the same suction (39 MPa) are similar, showing good repeatability.

The accuracy related to the deformation of the cell and the tubing can be evaluated indirectly through the experimental results. The results of heating tests showed that a dilation behavior can be identified in spite of the small difference between the test curve and the calibration one. In addition, the obtained dilation coefficient ($2 \times 10^{-4}$ °C$^{-1}$) is similar to that obtained by Romero (2005) on unsaturated compacted expansive clay. This shows that the cell has a satisfactory accuracy in terms of thermal volume changes. As far as the mechanical loading is concerned, since the mechanical volume change is larger than thermal volume change, the accuracy can be expected to be better.

**Conclusions**

A suction-temperature controlled isotropic cell was developed. Soil suction is controlled using the vapor equilibrium technique; soil temperature is controlled by immersing the cell in a bath having temperature controlled by a thermostat pump. The isotropic pressure of the cell is controlled by a volume/pressure controller. The volume change of soil specimen is estimated by the water volume change of the controller.

The results from two heating tests showed a dilation behavior of the soil; the thermal dilation coefficient is $2 \times 10^{-4}$ °C$^{-1}$. The effect of temperature on the soil compressibility behavior was studied. The compression tests at 39 MPa suction and different temperatures (20°C and 60°C), from 0.1 MPa to 50 MPa pressure, have showed that the compression and swelling index do not change significantly with temperature; however, the preconsolidation pressure strongly depends on temperature changes; it decreased from 1.9 MPa to 0.8 MPa when temperature was increased from 20°C to 60°C.

The experimental set-up offers satisfactory accuracy. Firstly, the volume/pressure controllers are accurate enough for pressure application and volume change measurements. Secondly, suction control was ensured by having careful control of temperature, which was achieved by immersing the whole cell in a thermostatically-controlled bath during testing. Although the volume measurement strongly depends on the deformation of the cell and the tubing, the calibration allows the measurements of thermal dilation coefficient and temperature effects on soil compressibility properties with satisfaction. In addition, the results obtained on two compression tests under 20 °C temperature are similar showing good repeatability of the system.

The developed cell is novel because it permits to perform thermo-mechanical tests at high suction (until 500 MPa), high temperature (20 – 80 °C) and high pressure (until 64 MPa). Therefore, this testing system can be used to study the thermo-hydro-mechanical behavior of heavily compacted expansive clays used in deep nuclear waste disposals. To the authors' knowledge, there have been no other testing devices with similar capacities allowing the characterization of clay barriers in this field.

**Acknowledgement**

The authors are grateful to French Electricity Company (EDF) for its financial support.

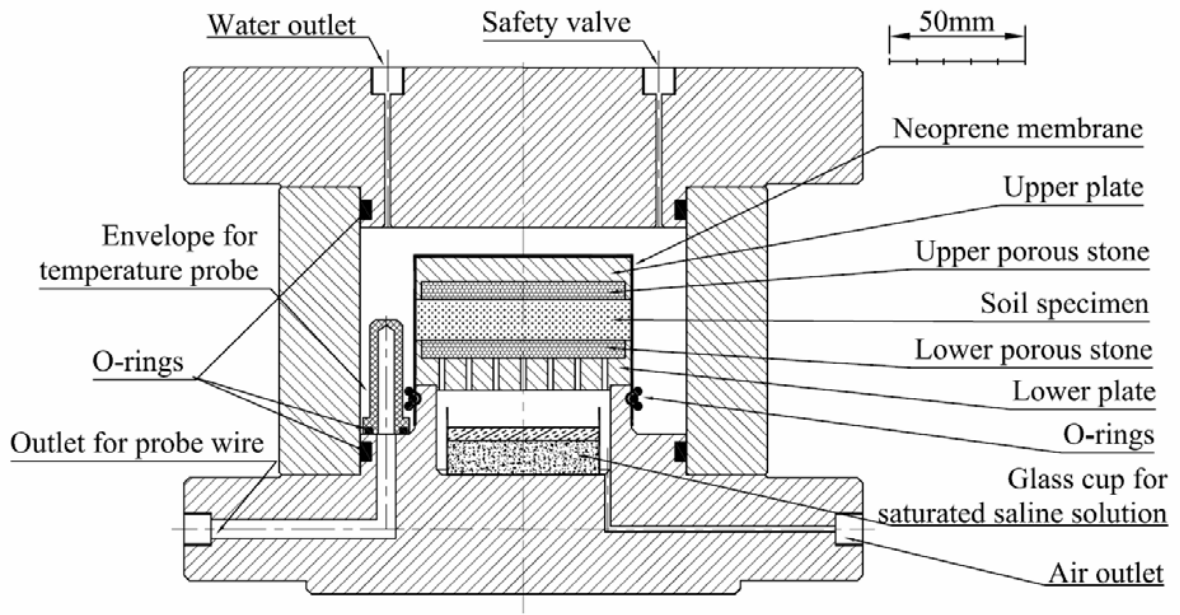

**Figure 1. Basic scheme of the suction-temperature controlled isotropic cell.**

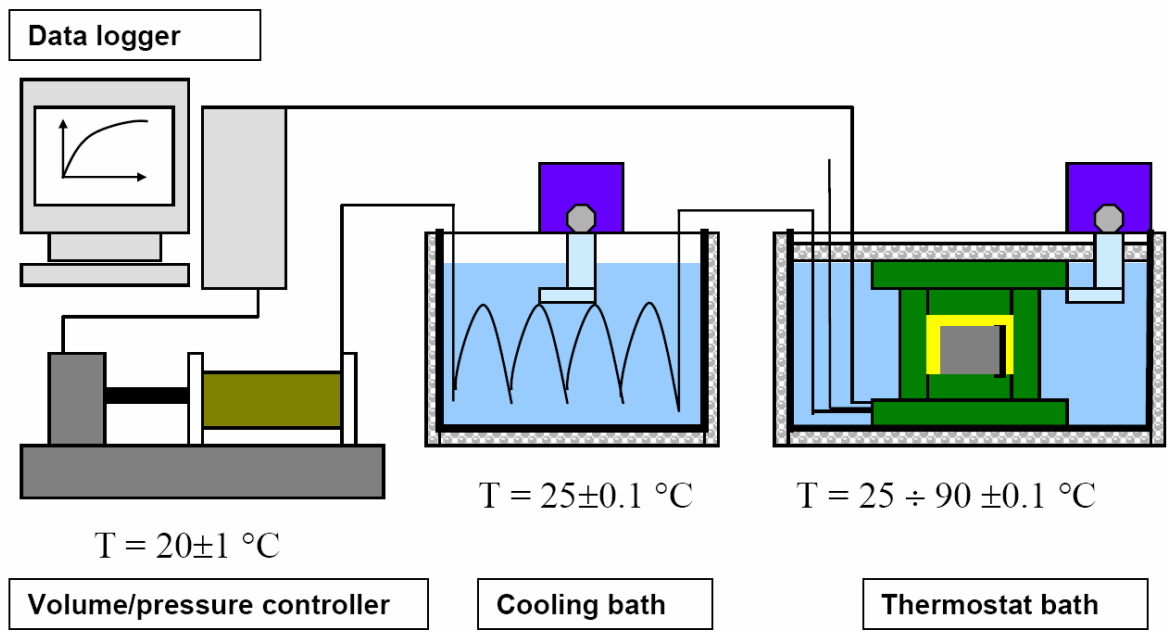

**Data logger**

T = 20±1 °C

T = 25±0.1 °C

T = 25 ÷ 90 ±0.1 °C

**Volume/pressure controller**   **Cooling bath**   **Thermostat bath**

**Figure 2. Experimental set-up for suction-temperature controlled isotropic compression test.**

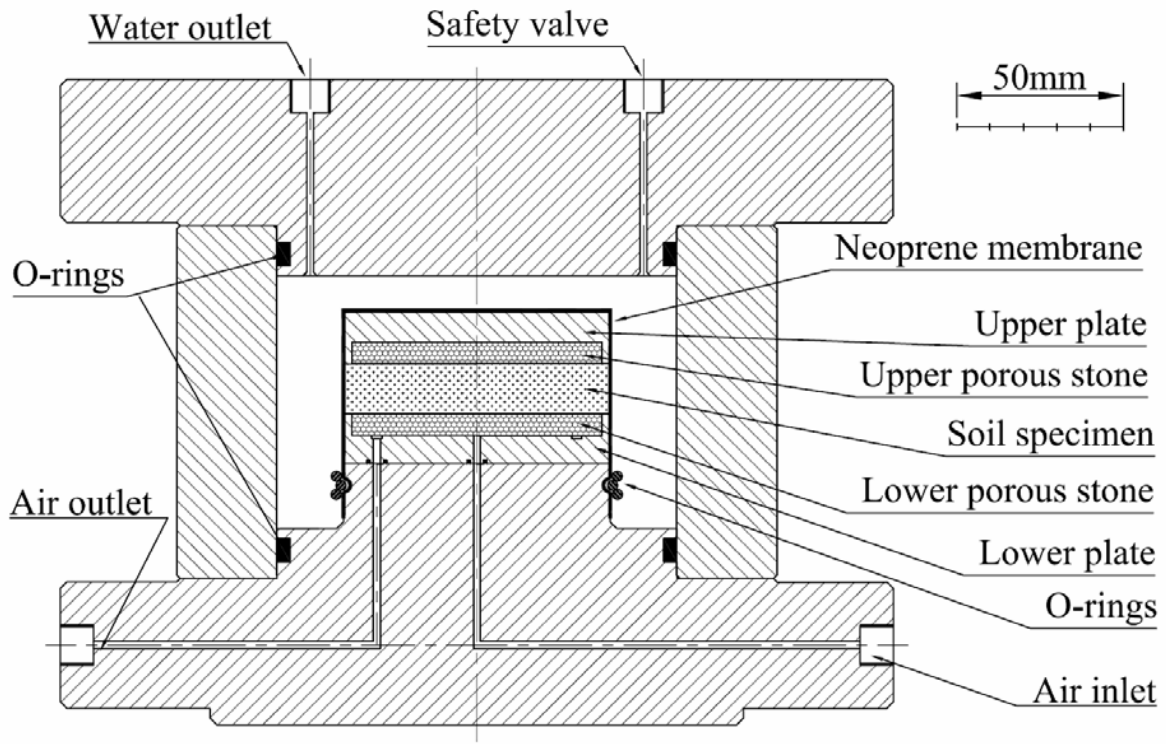

**Figure 3. Basic scheme of the isotropic cell with suction control by circulation of vapor.**

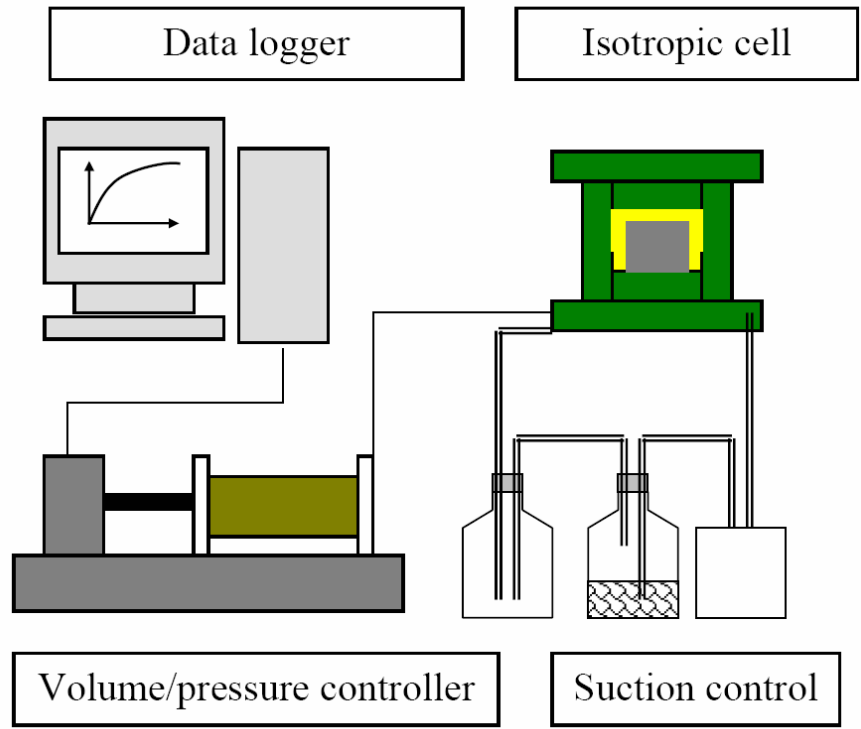

**Figure 4. Experimental set-up of suction controlled isotropic compression test with vapor circulation.**

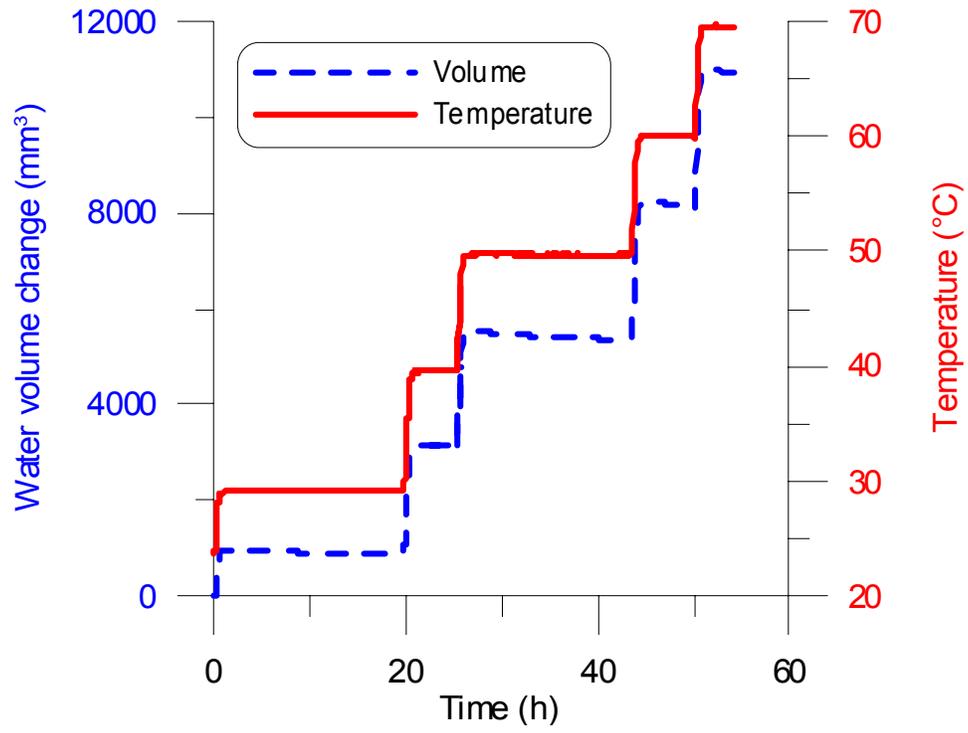

**Figure 5. Test T01. Water volume change and temperature versus time during heating.**

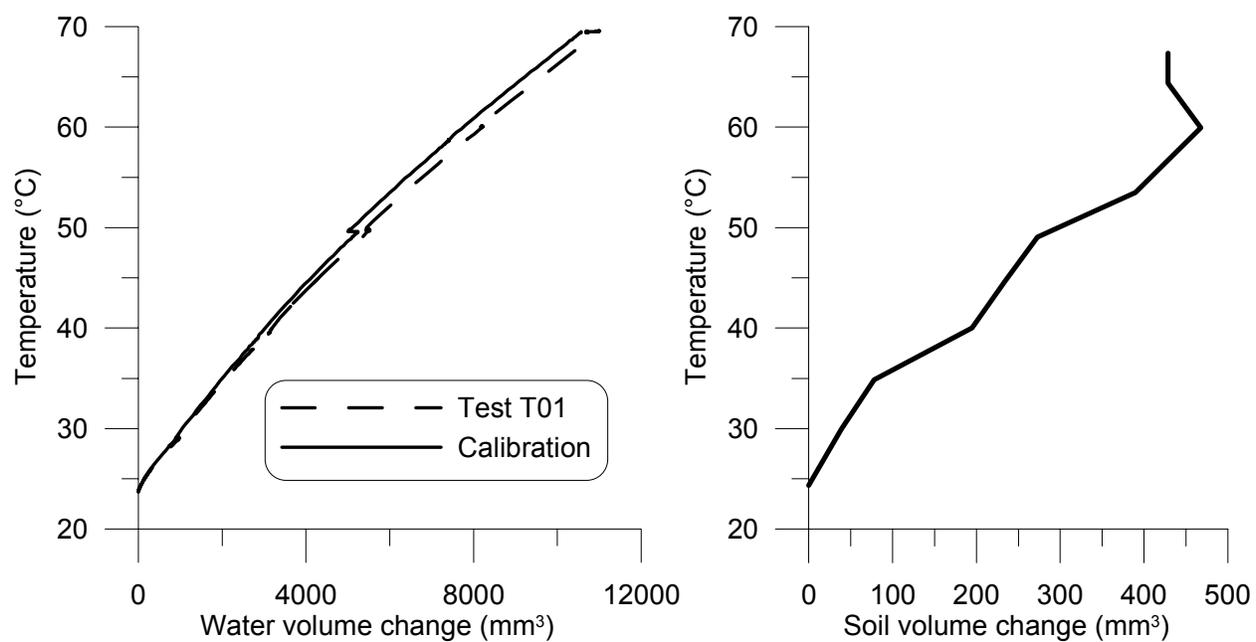

**Figure 6. Test T01. Water volume change and soil volume change versus temperature during heating.**

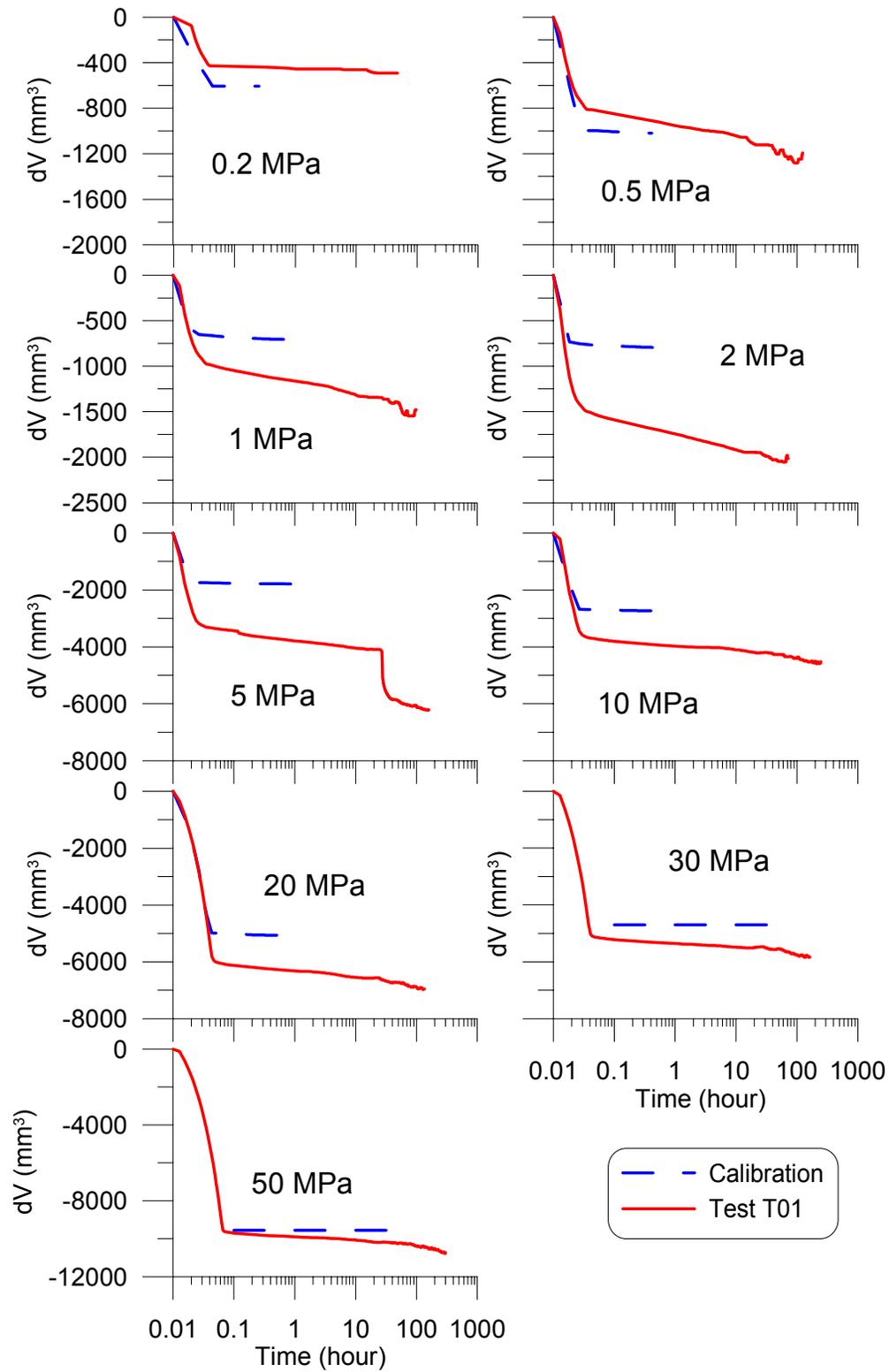

**Figure 7. Test T01. Water volume change versus time during loading steps.**

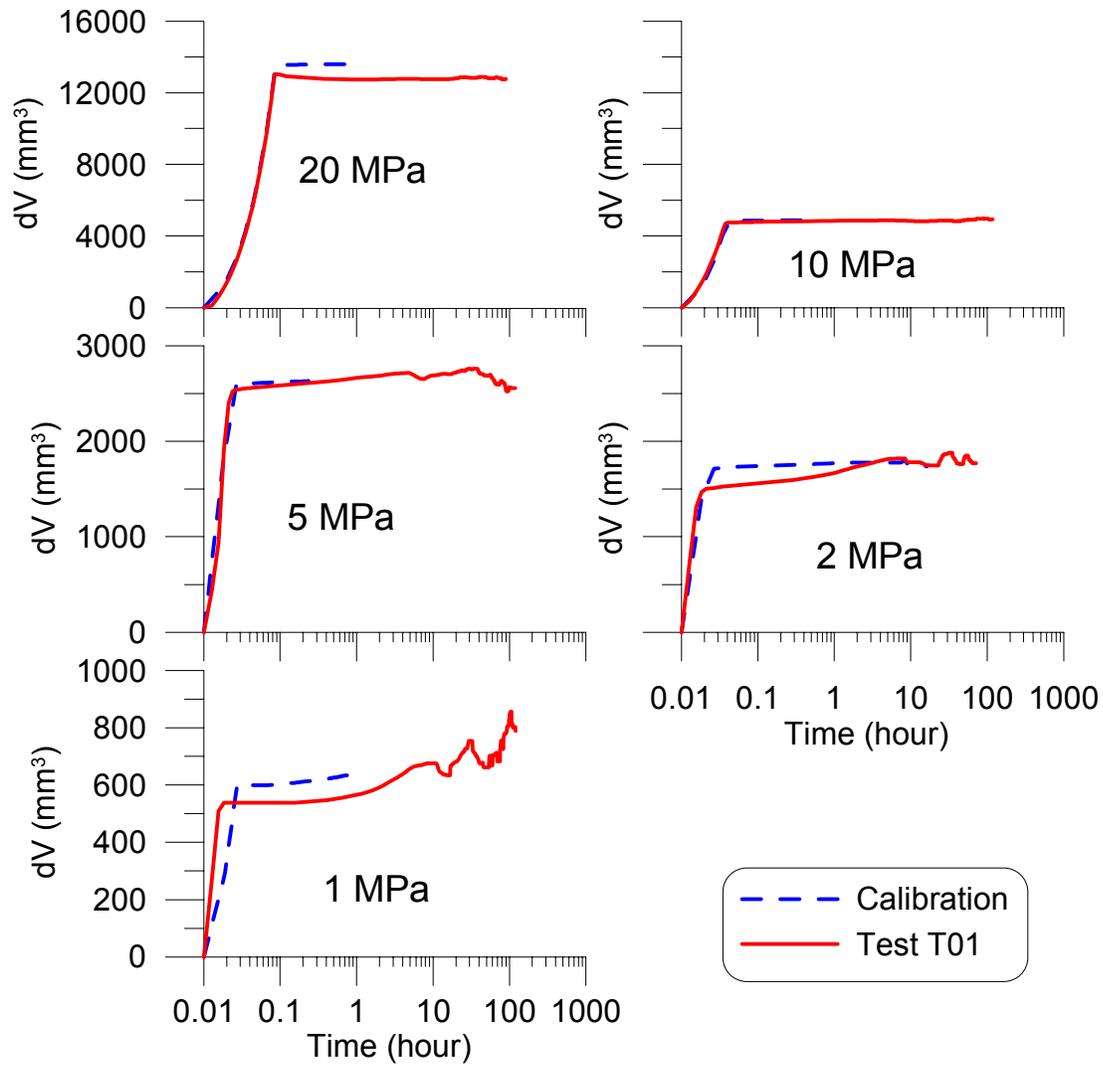

**Figure 8. Test T01. Water volume change versus time during unloading steps.**

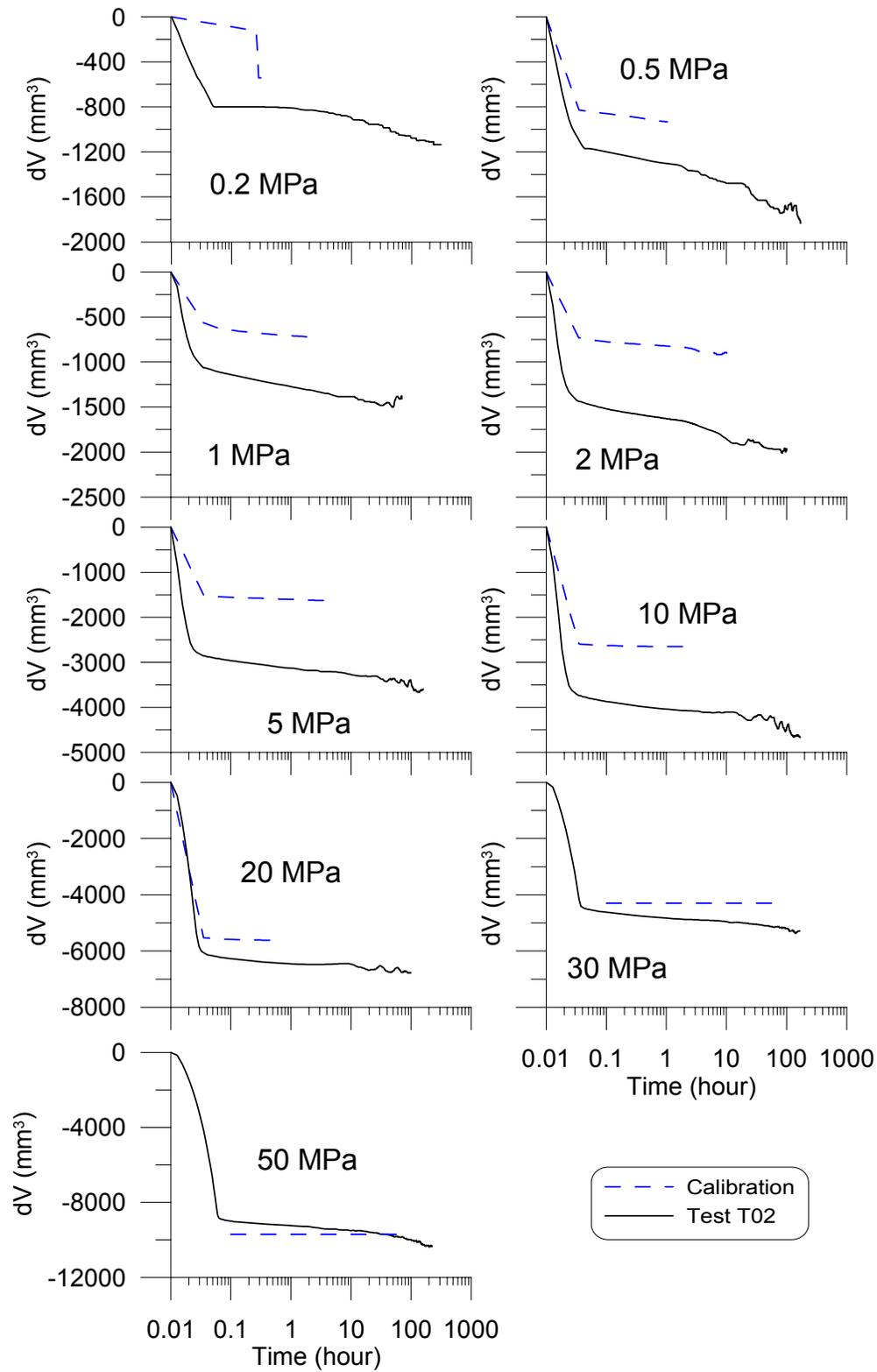

**Figure 9. Test T02. Water volume change versus time during loading steps.**

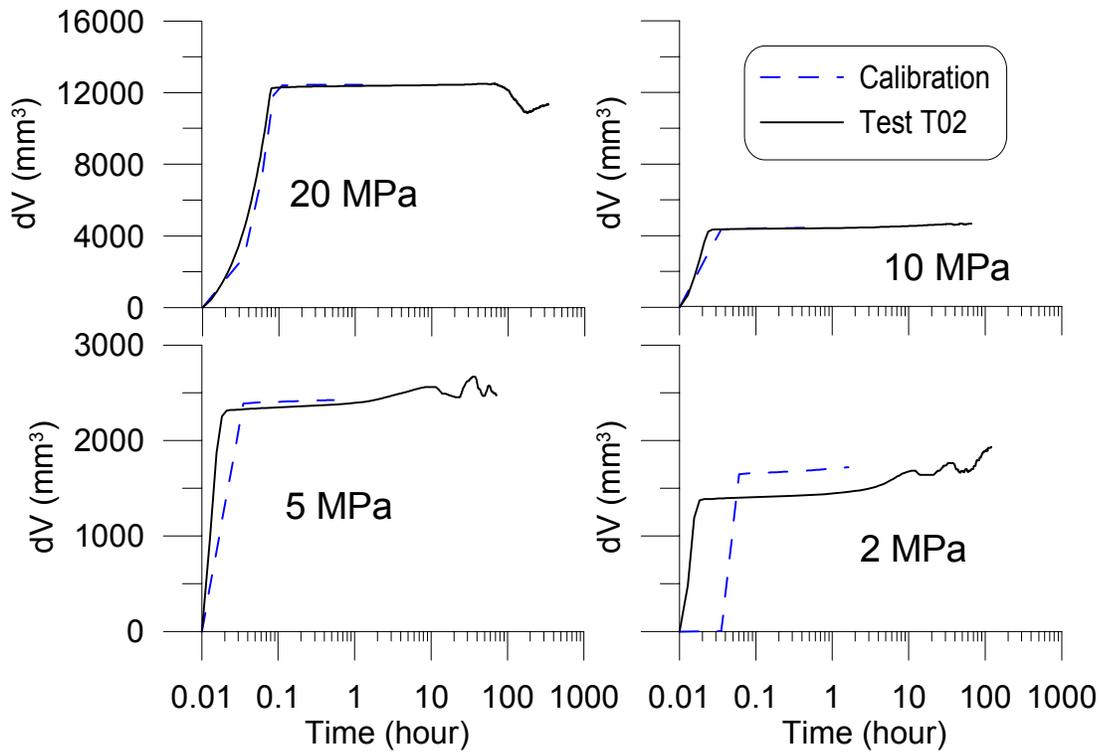

**Figure 10. Test T02. Water volume change versus time during unloading steps.**

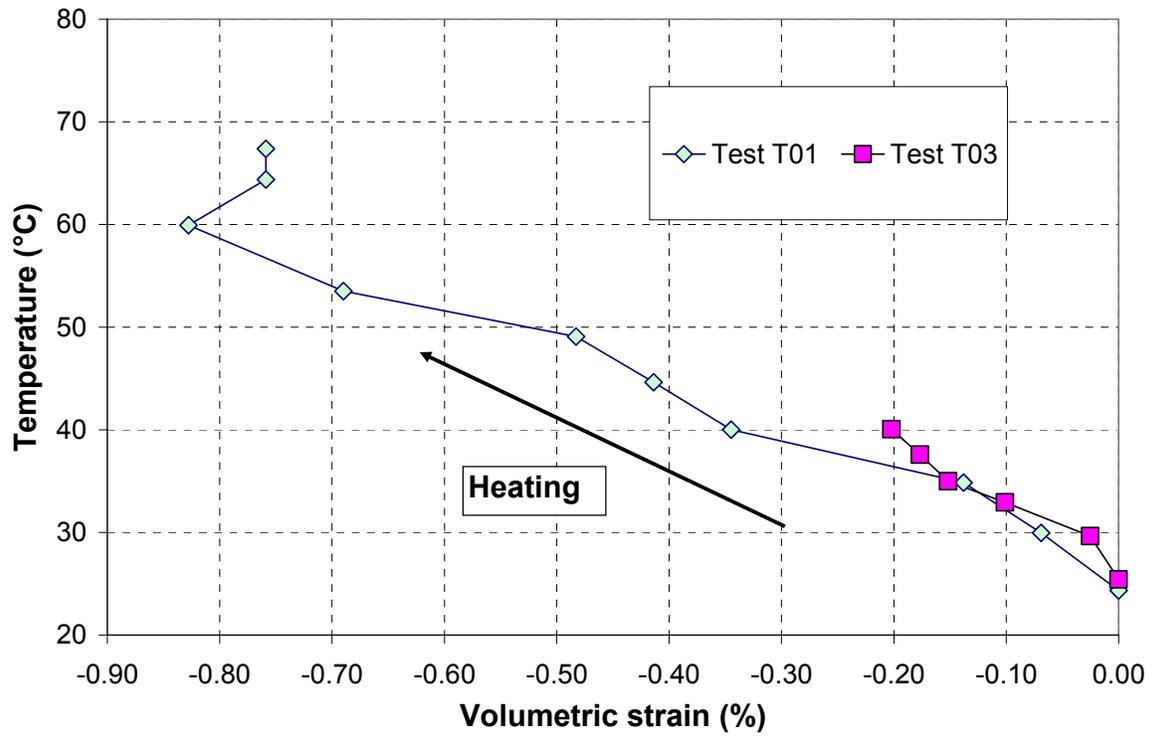

**Figure 11. Tests T01 et T03. Thermal volumetric strain versus temperature during heating.**

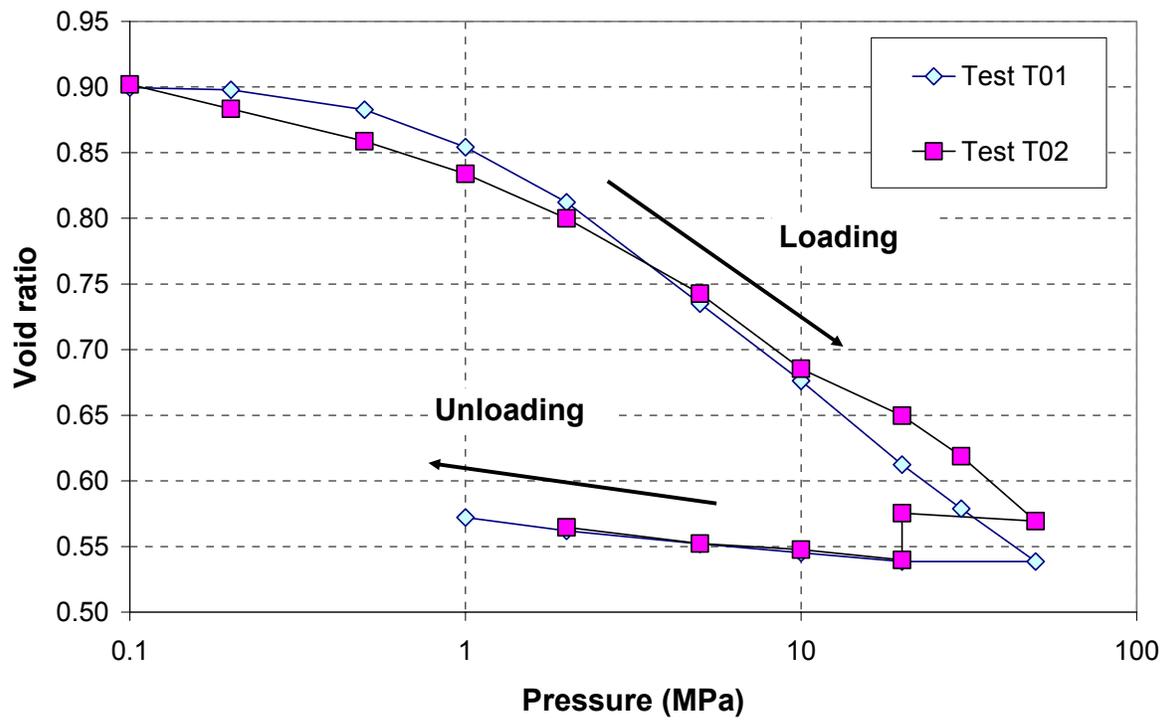

**Figure 12. Tests T01 and T02. Void ratio versus pressure during loading/unloading cycle.**

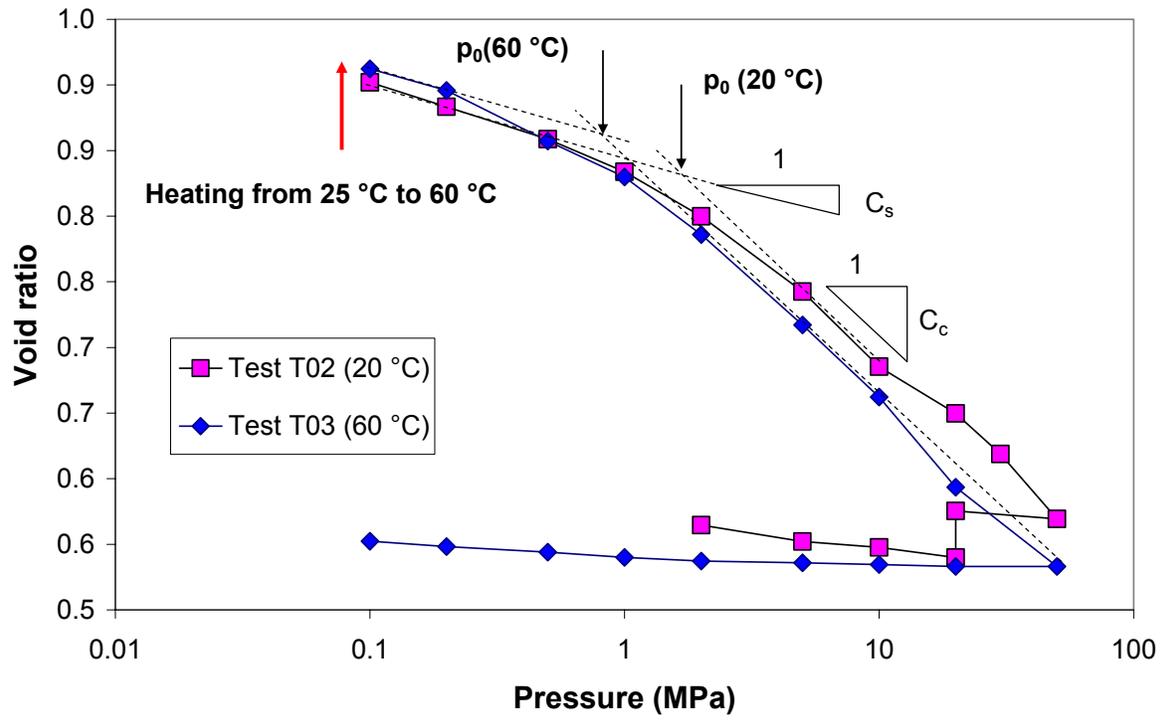

**Figure 13. Tests T02 and T03. Void ratio versus pressure during loading/unloading cycle.**